\documentclass[10pt]{iopart}
\expandafter\let\csname equation*\endcsname=\relax 
\expandafter\let\csname endequation*\endcsname=\relax 
\providecommand{\keywords}[1]{ #1}
\makeatletter
\def\ps@pprintTitle{%
 \let\@oddhead\@empty
 \let\@evenhead\@empty
 \def\@oddfoot{}%
 \let\@evenfoot\@oddfoot}
\makeatother
\renewcommand{\thetable}{\Roman{table}}
\usepackage{csquotes}
\usepackage{graphicx,color}
\usepackage{float}
\usepackage{amsmath}
\usepackage{amssymb}
\usepackage{multirow}
\usepackage{dsfont}
\usepackage{adjustbox}
\usepackage{pdflscape}
\usepackage{wasysym}
\usepackage[percent]{overpic}
\usepackage[dvipsnames]{xcolor}
\usepackage[normalem]{ulem}
\usepackage{subcaption}
\begin{document}

\title{Runaway electron induced explosions of graphite; modeling versus controlled DIII-D experiments}
\author{T. Rizzi$^{a}$, A. Kavroulakis$^{b,c}$, S. Ratynskaia$^{a}$, P. Tolias$^{a}$, M. Tatarakis$^{b,c}$, E. Kaselouris$^{b,d}$, V. Dimitriou$^{b,d}$, E. Hollmann$^{e}$, F. Brochard$^{f}$, A. Kulachenko$^{g}$,  M. Beidler$^{h}$, C. Lasnier$^{i}$, R. A. Pitts$^{j}$}
\address{$^a$\,Electromagnetics and Plasma Physics, KTH Royal Institute of Technology, Stockholm SE-100 44, Sweden \\
$^b$\,Institute of Plasma Physics and Lasers-IPPL, University Research and Innovation Centre, Hellenic Mediterranean University, 74150 Rethymnon, Greece \\
$^c$\,Department of Electronic Engineering, Hellenic Mediterranean University, 73133 Chania, Greece\\
$^d$\,Department of Music Technology and Acoustics, Hellenic Mediterranean University, 74133 Rethymnon, Greece\\
$^e$\,University of California - San Diego, La Jolla, CA, USA \\
$^f$\,Institut Jean Lamour, Universite de Lorraine, CNRS, F-54000 Nancy, France \\
$^g$\,Mechanics, KTH Royal Institute of Technology, Stockholm SE-100 44, Sweden \\
$^h$\,Oak Ridge National Lab, Oak Ridge, TN, USA \\
$^i$\, Lawrence Livermore National Lab, Livermore, CA, USA\\
$^j$\,ITER Organization, Saint-Paul-lez-Durance, France}

\begin{abstract}
\noindent 
The state-of-the-art concerning the modeling of the thermomechanical response of graphite to runaway electron (RE) impact is based on one-way coupled linear thermoelasticity combined with Rankine's strength-based failure criterion and limited to the onset of material failure. Here, the predictive capabilities are extended to the nonlinear damage phase characterized by material fragmentation and debris expulsion. This is achieved by introducing plasticity via the Johnson-Holmquist constitutive model, adopting an effective plastic strain-based failure criterion and coupling finite element analysis with smoothed-particle hydrodynamics. The extended thermomechanical model is successfully benchmarked against the results of two controlled RE-induced damage experiments recently carried out in DIII-D. This constitutes an important step towards the final objective of quantitatively describing the thermomechanical response of tungsten to REs.

\end{abstract}
\maketitle
\ioptwocol

\section{Introduction}\label{sec:introduction}

In the technologically mature tokamak magnetic confinement concept, provision of plasma-facing components (PFC) with a sufficient lifetime constitutes one of the major hurdles in the development of commercial nuclear fusion reactors\,\cite{Pitts_2025,Pitts2019,Krieger_2025,Ratynskaia_2025b}. A critical aspect of PFC damage concerns the uncontrolled incidence of relativistic runaway electrons (REs), which locally generates extreme volumetric power densities. Instances of accidental damage from major tokamaks world-wide have shown that RE impacts often result in an explosive PFC response accompanied by ejection of debris\,\cite{Ratynskaia_2025b}. This has been confirmed in controlled experiments with graphite samples in DIII-D\,\cite{Hollmann2025,Hollmann_2025b} and a tungsten (W) tile in WEST\,\cite{Corre_2025}. This expanding experimental dataset has played a pivotal role in advancing the understanding of the PFC thermo-mechanical behavior under RE loading. Noteworthy progress has been achieved in the modeling of the response of sublimating brittle PFCs, also enabled by the absence of a liquid phase and viscoplasticity\,\cite{Ratynskaia_2025}. The same thermomechanical model was successfully applied to describe the accidental damage of boron nitride tiles exposed to REs in WEST\,\cite{Rizzi_2026}.

In our previous work, a KORC-Geant4-COMSOL workflow was demonstrated to be capable of predicting the onset of brittle failure within one-way coupled linear thermo-elasticity and Rankine's failure theory\,\cite{Ratynskaia_2025,Rizzi_2026}. Here, the predictive capabilities are extended to the final nonlinear damage phase that is characterized by material fragmentation and debris ejection. This is achieved by employing the Johnson-Holmquist damage model and combining finite element analysis\,\cite{Kaselouris_2017_NatCom,Kaselouris_2021_PPCF} with a smooth particle hydrodynamics approach in LS-DYNA\,\cite{Hallquist_2006}. The full thermomechanical model is benchmarked against the results of the 2023 and 2024 DIII-D experiments\,\cite{Hollmann2025,Hollmann_2025b}.  Validation of the complete thermomechanical response of graphite to REs marks a major milestone on the way to the final objective to develop a simulation framework for the assessment of RE-induced damage on W, the PFC material of choice for most tokamak reactors, including ITER\,\cite{Pitts_2025}.

\section{Experimental evidence and input}\label{sec:experimental}

\begin{table*}[h!]
\centering
\setlength{\tabcolsep}{4pt}
\begin{tabular}{c c c c c c c c c c c}
\hline
{Year} &  \begin{tabular}{c}{Shot}   \\ {\#}\end{tabular} & \begin{tabular}{c}{Energy}   \\ {deposited}\end{tabular}  &\begin{tabular}{c}{Loading}   \\ {duration}\end{tabular}  & \begin{tabular}{c}{Erosion}   \\ {depth}\end{tabular}  & \begin{tabular}{c}{Eroded}   \\ {surface}\end{tabular} &\begin{tabular}{c}{Blown-off}   \\ {volume }\end{tabular} & {Debris} & \begin{tabular}{c}{RE parameters}   \\ {at impact}\end{tabular}    &  Modeled\\ 
\hline
2023 & 191366 & 10 kJ  & 1 ms & 1.2 mm & 100 mm$^2$ & 140 mm$^3$ & Yes & 3 MeV 25$^\circ$& Here, Case\,1\\
2024 & 200236 & 0.7 kJ & 1 ms & N/A & N/A & N/A & No & 4 MeV 10$^\circ$&  Ref.\cite{Hollmann_2025b}\\
2024 & 200237 & 1.4 kJ & 1 ms & N/A & N/A & N/A & No & 4 MeV 10$^\circ$&  Ref.\cite{Hollmann_2025b}\\
2024 & 200241 & 9.8 kJ & 1 ms & 0.9 mm & 50 mm$^2$ & 30 mm$^3$   & Yes & 4 MeV 10$^\circ$& Here, Case\,2\\
\hline
\end{tabular}
\caption{Summary of the RE impact, RE loading  and PFC damage parameters in the two experimental campaigns.}
\label{tab:summary_exposure}
\end{table*}

Instrumented sacrificial limiters made of ATJ graphite were employed in DIII-D as targets for RE impact during low Z terminations in dedicated discharges executed in the 2023 and 2024 experimental campaigns\,\cite{Hollmann2025, Hollmann_2025b}. The limiters, located in the lower divertor, were equipped with different diagnostics, allowing a first-of-a-kind modeling of RE final loss instabilities. The RE kinetic energy and pitch angle were empirically reconstructed\,\cite{Hollmann2025} and supported by numerical simulations with KORC (Kinetic Orbit Runaway electrons Code)\,\cite{Castillo_2018}. The inclusion of novel thermoluminescent detectors (TLDs) in the instrumented DIII-D limiter allowed further investigation of the impact dynamics and local structure of the RE beam\,\cite{Hollmann_2025b}. 

The sample geometry was identical in both DIII-D campaigns and can be found in Fig.1 of Ref.\cite{Ratynskaia_2025}. A calibrated fast thermocouple (TC) was inserted $\sim$2\,cm below the top (plasma-facing) surface to measure the total energy deposited. Two IR cameras monitored the limiters during exposures: a close camera measuring the spatial heat load distribution at the surface, and a distant camera capturing the dynamics of the debris. Moreover, hard X-ray (HXR) detectors were employed for estimates of the energy deposition time.

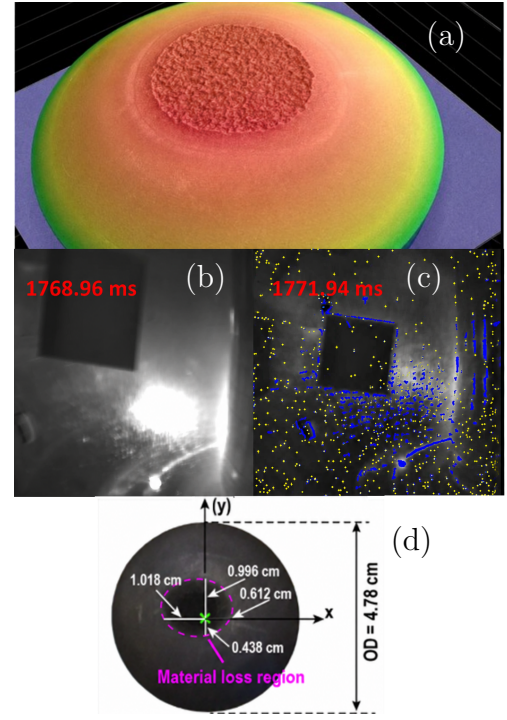
\begin{figure}[!h]
    \centering
    \begin{overpic}[width=0.78\linewidth]{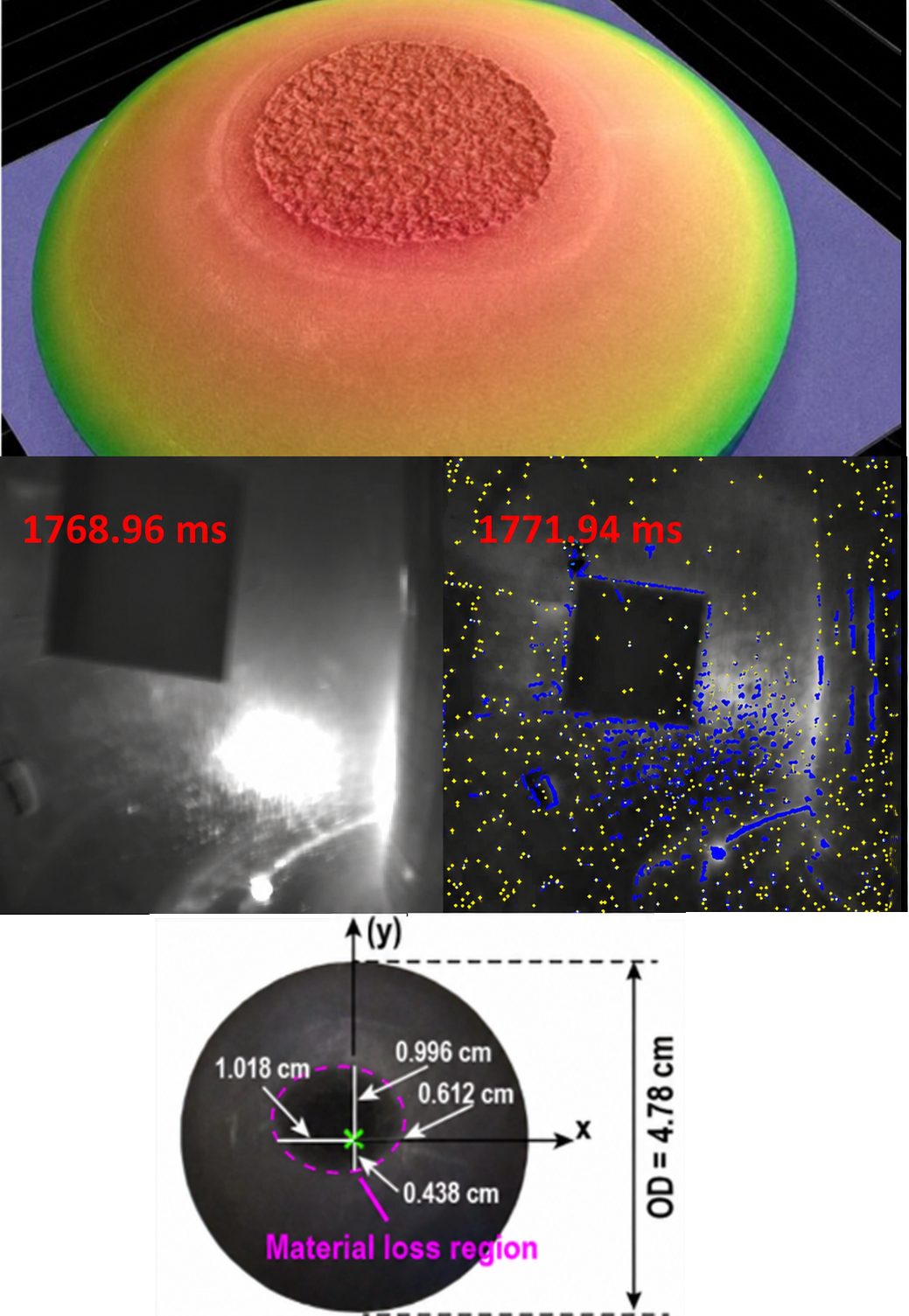}
        \put(58,94){\large \textcolor{white}{(a)}}
        \put(24,60){\large  \textcolor{white}{(b)}}
        \put(55,60){\large  \textcolor{white}{(c)}}
        \put(53,23){\large (d)}
    \end{overpic}
    \caption{Case 1 (\#191366): (a) post-mortem confocal image of damaged sample (reprinted from Ref.\cite{Ratynskaia_2025}); (b) snapshot of debris cloud at $1768.96$\,ms (RE impact at 1765.96), (c) snapshot of debris cloud at $1771.94$\,ms with detected fragments marked in yellow and blue (see the text for the color coding).  Case 2 (\#200241): (d) post-mortem image of the sample also highlighting the material loss region.}
    \label{fig:geom}
\end{figure}

The parameters that serve as input or are required for the benchmarking are summarized in Table\,\ref{tab:summary_exposure}, where columns 2-7 compile key experimentally deduced in-situ and post-mortem quantities, while column 8 reports RE impact parameters as reconstructed by the modeling. Since the goal is to validate the expanded work-flow that includes material fragmentation and debris expulsion, only exposures resulting in explosions will be considered, hereafter referred to as Case 1 (2023) and Case 2 (2024). An overview of the material damage is provided in Fig.\ref{fig:geom}.


The IR camera used for debris observations had $\sim1$\,ms sample rate. High debris concentration combined with the low acquisition rate prevents resolution of the individual motion for the majority of fragments. The task is complicated by the strong thermal radiation from the debris cloud and the surroundings. Despite these limitations, the Particle Tracking Velocimetry (PTV) software AX R\&D \cite{Helle_2024} was able to identify close to 1000 particles per frame by using adaptive histogram optimization and auto-adaptative multiscale thresholding\,\cite{Brochard_2017}. 

In Case 1, Fig.\ref{fig:geom}(c) shows an example of fragment detection in a later frame, when the cloud is disintegrating, compare (b) with (c). Those marked as yellow are interpreted as individual fragments/particles, while those marked as blue are considered too large and interpreted as agglomerates/not resolved and thus are excluded from tracking analysis. The majority of the debris have speeds of a few tens of m/s, while a small, $\sim 10 \%$, population exhibits speeds in the range of 200-400\,m/s. The speed distribution of individual fragments obtained with a Kalman filter is shown in the comparison with the simulation results, see Sec.\ref{sec:results1}. In addition to the $<10$\,m/s instrumental error, there are association errors (mismatches) due to the high fragment concentration, which are estimated by tracking simulated data that replicate the experimental video yielding $\sim 20\%$ for the particle number per speed distribution bin. For particles with speeds $>100$ m/s, association errors are difficult to quantify. Note that results obtained from 2D tracking are likely to be lower than the actual 3D speeds. The Case 2 video is of much lower quality; with a few tens of particles identified, thus the reconstructed distribution is highly uncertain. 

\section{Modeling approach and methodologies} \label{sec:modeling}

\subsection{Energy deposition}

The RE energy deposition is evaluated by Monte Carlo simulations with Geant4, described in detail in Ref.\cite{Ratynskaia_2025}. The RE impact parameters constitute external input for Geant4, which are deduced by different means. 

For the 2023 campaign, the RE impact parameters were reconstructed by KORC, which tracks full-orbit relativistic electron trajectories in 3D electromagnetic fields\,\cite{Carbajal_2017,Beidler_2020}. In the modeling of Case 1, the 3\,MeV \& 25$^\circ$ pitch scenario will be followed, since it leads to accurate predictions for the onset of brittle failure\,\cite{Ratynskaia_2025}. 

For the 2024 campaign, the RE impact parameters were reconstructed assuming mono-energetic, mono-pitch distributions\,\cite{Hollmann_2025b}, with the magnetic field obtained via current filament reconstruction from JFIT\,\cite{Humphreys_1999}, neglecting MHD perturbations. In the modeling of Case 2, the 4\,MeV \& 10$^\circ$ pitch scenario will be followed, since it provides the best match with the TLD data\,\cite{Hollmann_2025b}. 

As shown in Table \ref{tab:summary_exposure},  Cases 1 and 2 differ significantly in terms of the blown-off volume, despite the similar initial RE parameters and energy deposited. This is mainly due to variations in the RE beam wetted area; In Case 2, the REs are spread over a wider area, resulting in weaker in-depth energy density gradients and $\sim20\,\%$ lower peak energy density values.

\subsection{Thermomechanical model}

Material explosions are driven by thermal expansion, thus requiring the coupling of thermal and mechanical responses. The temperature field evolves according to
\begin{equation}
\rho{c}_{\mathrm{p}}\frac{\partial T}{\partial t}=\nabla\cdot(k\nabla{T})+Q
\end{equation}
where $\rho$ is the mass density, $c_{\mathrm{p}}$ is the specific isobaric heat capacity, $k$ is the thermal conductivity and $Q$ the volumetric heat generation rate (W/m$^3$). The latter is retrieved from the Geant4 simulations\,\cite{Ratynskaia_2025}.

In a typical Lagrangian formulation, the governing equations for the mechanical response include the continuity and momentum equations, which read as
\begin{equation}
    \frac{d\rho}{dt} + \rho\nabla\cdot\mathbf{u} = 0\,
    \label{eq:mass}
\end{equation}
\begin{equation}
\rho\frac{\partial \mathbf{u}}{\partial t}=\nabla\cdot\boldsymbol{\sigma}\,,
\label{eq:energy}
\end{equation}
in absence of body forces, with $\boldsymbol{\sigma}$ the stress tensor and $\mathbf{u}$ the velocity field. The thermomechanical coupling is realized mainly via inclusion of the thermal strain $\epsilon_{\mathrm{T}} =\alpha_{\mathrm{T}}(T-T_0)$, where $\alpha_{\mathrm{T}}$ is the thermal expansion coefficient and $T_0$ the initial temperature, but also via the temperature-dependent mechanical properties. These equations are valid in the finite element method (FEM) and smoothed-particle hydrodynamic (SPH) formulation\,\cite{Monaghan_2005}, but require a closure equation that varies depending on the material model.

\subsection{Material constitutive model}

In our previous work\,\cite{Ratynskaia_2025}, the linear thermoelastic constitutive relation for isotropic materials was employed to describe the behavior of ATJ graphite up to the onset of failure. The general expression reads as
\begin{equation}
 \boldsymbol{\sigma} = 2\mu\boldsymbol{\epsilon} +\lambda\text{Tr}(\boldsymbol{\epsilon})\mathbf{I} - (2\mu + 3\lambda)\alpha_{\mathrm{T}}(T-T_0)\mathbf{I} \,.
    \label{eq:constitutive_lin_el}
\end{equation}
Here, $\boldsymbol{\epsilon}$ is the strain tensor, while $\lambda$,\,$\mu$ are the first and second Lamé parameters. The Rankine criterion was utilized, according to which local principal stresses are compared with the temperature-dependent ultimate tensile or compressive strengths, depending on their sign, and failure occurs if either $\sigma_1 > \mathrm{UTS}(T)$ or $|\sigma_3|>\mathrm{UCS}(T)$. This model was able to successfully identify the onset of failure\,\cite{Ratynskaia_2025}. However, a linear thermoelastic description cannot capture material fragmentation and damage evolution. 

In the present work, the Johnson-Holmquist 2 (JH-2)\,\cite{Johnson_1994} constitutive model is adopted and coupled with a strain-based failure criterion. The JH-2 is a pressure, strain-rate and damage dependent model, specifically formulated to describe brittle materials that are subject to large strains, high strain rates, and high pressures, by blending elasto-plastic behavior with progressive damage and fracture. In fact, it allows for plastic deformation, even though the material is brittle: the material damage accumulates as a function of the effective plastic strain to fracture, which depends on pressure. As damage grows, the material progressively softens while still supporting finite strength. In JH-2, the equivalent Von Mises stress ($\sigma_{\mathrm{eq}})$ is explicitly defined as a continuous interpolation between an intact strength surface $\sigma_i$ and a fracture strength surface $\sigma_{\mathrm{f}}$, weighted by a scalar damage variable $D\in[0,1]$, so that the material response transitions smoothly rather than abruptly from intact to fully fractured states. However, even when the scalar damage reaches unity, this does not imply direct element removal, since the element can still withstand hydrostatic pressure. In fact, the element is flagged as \emph{failed} when its effective plastic strain (EPS) overcomes a given limit, which is $\epsilon_{\mathrm{fail}}=0.025$ for typical ATJ graphite\,\cite{Kachur_1964}.

The constitutive equations of the JH-2 model are 
\begin{align}
\sigma^{*} &= (1-D)\,\sigma_i^{*} + D\,\sigma_f^{*}\,,\nonumber\\
\sigma_i^{*} &= A\left(P^{*} + P_{\mathrm{T}}^{*}\right)^{n}\left( 1 + C \ln \dot{\epsilon}^{*} \right)\,, \\
\sigma_f^{*} &= B (P^{*})^{m}\left( 1 + C \ln \dot{\epsilon}^{*} \right) \nonumber\,,
\end{align}
where $A$,$B$,$C$,$n$,$m$ are characteristic strength parameters of the material, $P$ and $P_{\mathrm{T}}$ are the pressure and the maximum tensile pressure respectively, $\dot{\epsilon}$ is the strain rate and the quantities that feature a $^*$ superscript are normalized with respect to their values at the Hugoniot elastic limit. The above constitutive equations are coupled to a polynomial equation of state that governs the hydrostatic behavior\,\cite{Johnson_1994}. All parameters employed in the implementation of the JH-2 material model for ATJ graphite are obtained from Refs.\,\cite{Huddleston_2023,Jaulin_2018}. In this formulation, the thermal expansion is introduced separately, since the constitutive model itself does not inherently couple the temperature to the volumetric strain. Hence, the thermal strain $\epsilon_{\mathrm{T}}$ is computed at each time step and subtracted from the total strain.


\subsection{Adaptive FEM-to-SPH simulations of fragmentation}

Purely FEM-based descriptions are not sufficient to capture fragmentation and debris dynamics. In fact, element-deletion methods remove the failed element immediately after the failure criteria are satisfied. To overcome this obstacle, an adaptive conversion to the SPH formulation is adopted\,\cite{Monaghan_2005}. SPH is a mesh-free, Lagrangian numerical method in which a continuum is represented by a set of discrete particles that carry material properties such as mass, velocity, stress, and internal energy, and which evolve in time according to the governing conservation laws, see Eqs.(\ref{eq:mass}-\ref{eq:energy}) above. The field variables are not stored on a grid but are reconstructed locally by kernel-weighted interpolations over neighboring particles within a finite smoothing length, which allows spatial derivatives to be computed without a predefined mesh.         

The LS-DYNA software is employed being well suited for hybrid FEM–to-SPH simulations and mesh-free formulations. The adaptive model transforms any element of the Lagrangian continuum that violates the EPS-based failure criterion into an SPH particle. Newly generated SPH particles inherit the Lagrange nodal quantities of the failed elements and remain numerically coupled to the adjacent intact FEM elements. The heat equation is not solved for the SPH particles, whose temperature remains constant. Even if the internal energy grows (plastic work), no RE loading is received by SPH particles post generation.

The thermal problem features insulating boundary conditions, since radiative and vaporization cooling are negligible due to the moderate temperatures and short loading times. The mechanical problem assumes zero initial displacement, with the side and top surfaces free and clamped, and outflow (non-reflecting) boundary conditions for the bottom surface.


\section{Results for Case 1 (\#191366)}\label{sec:results1}

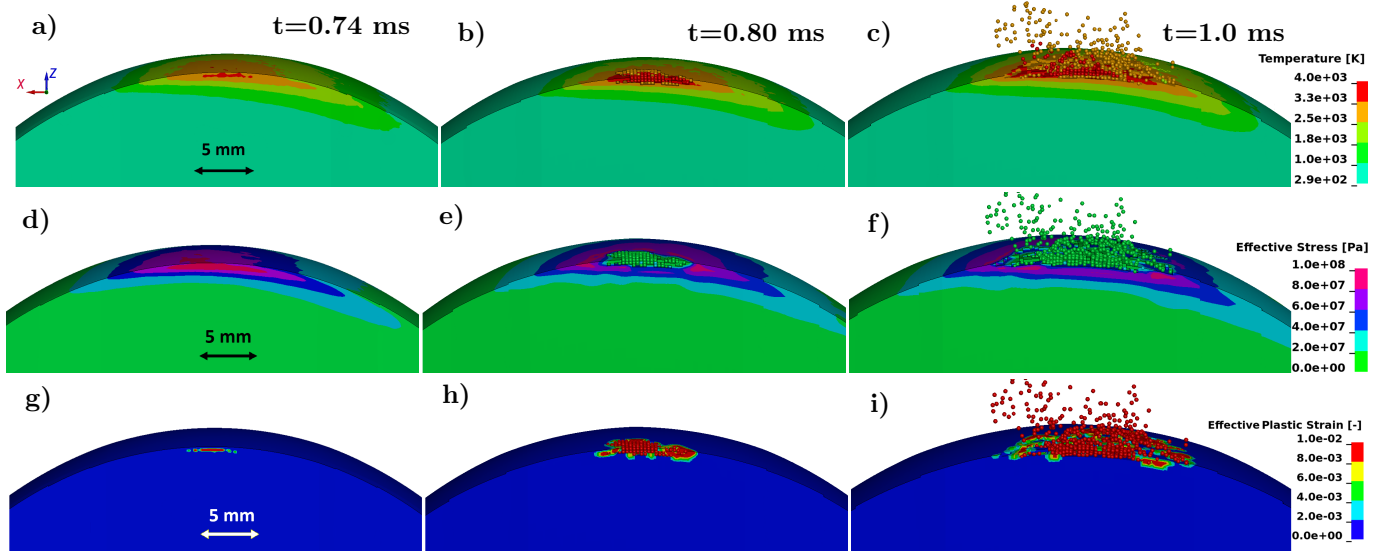
\begin{figure*}[!t]
    \centering
    \addtolength{\tabcolsep}{-9pt} 
    \setlength{\tabcolsep}{-2pt}
    \begin{tabular}{@{}c@{}c@{}c@{}}        
        \setlength{\tabcolsep}{-2pt}
        \subfloat{%
          \begin{overpic}[width = 2.2in]{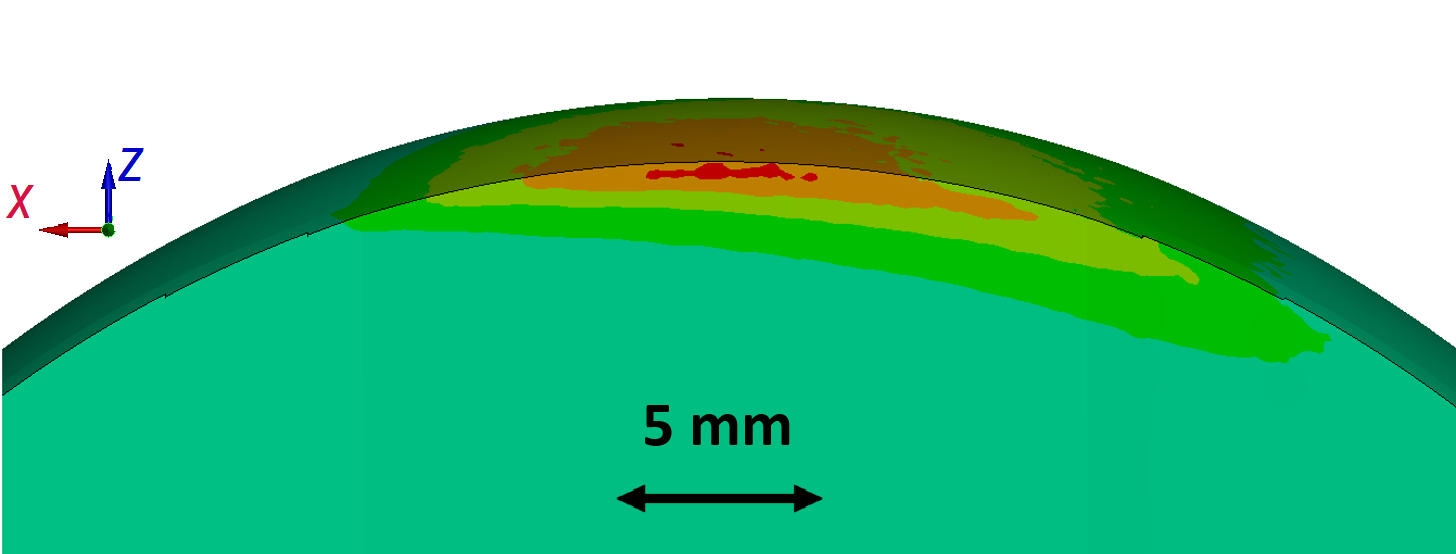}
          \put(2,36){\hbox{\kern3pt\textcolor{black}{\textbf{a)} }}}
          \put(59,36){\hbox{\kern3pt\textcolor{black}{\textbf{t=0.74 ms} }}}
          \end{overpic}
        }
        \hspace{-3mm}
        \subfloat{%
          \begin{overpic}[width = 2.1in]{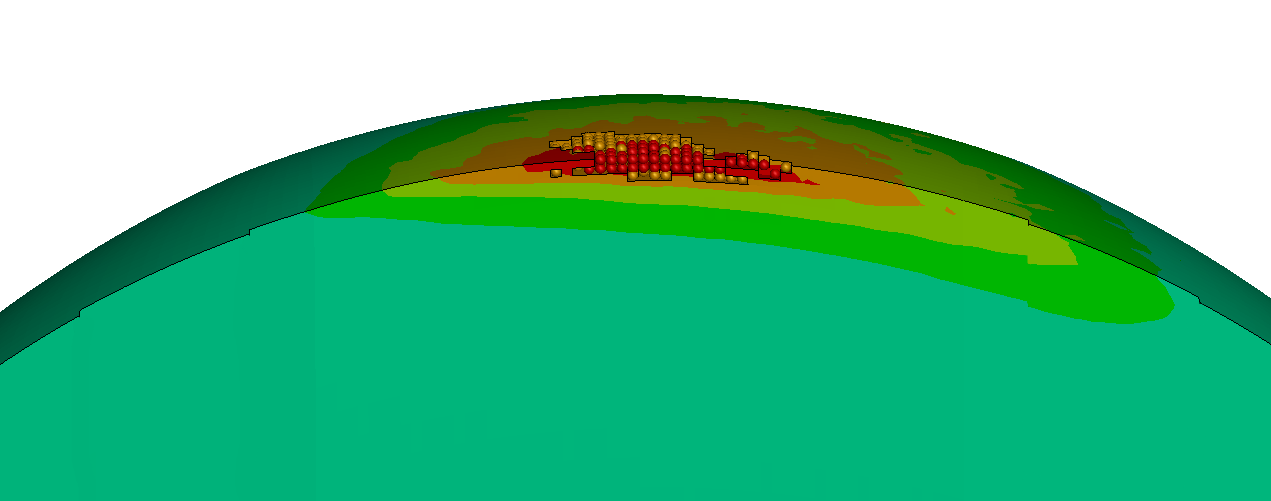}
          \put(2,36){\hbox{\kern3pt\textcolor{black}{\textbf{b)} }}}
          \put(59,36){\hbox{\kern3pt\textcolor{black}{\textbf{t=0.80 ms} }}}
          \end{overpic}
        }
        \hspace{-3mm}
        \subfloat{%
          \begin{overpic}[width = 2.7in]{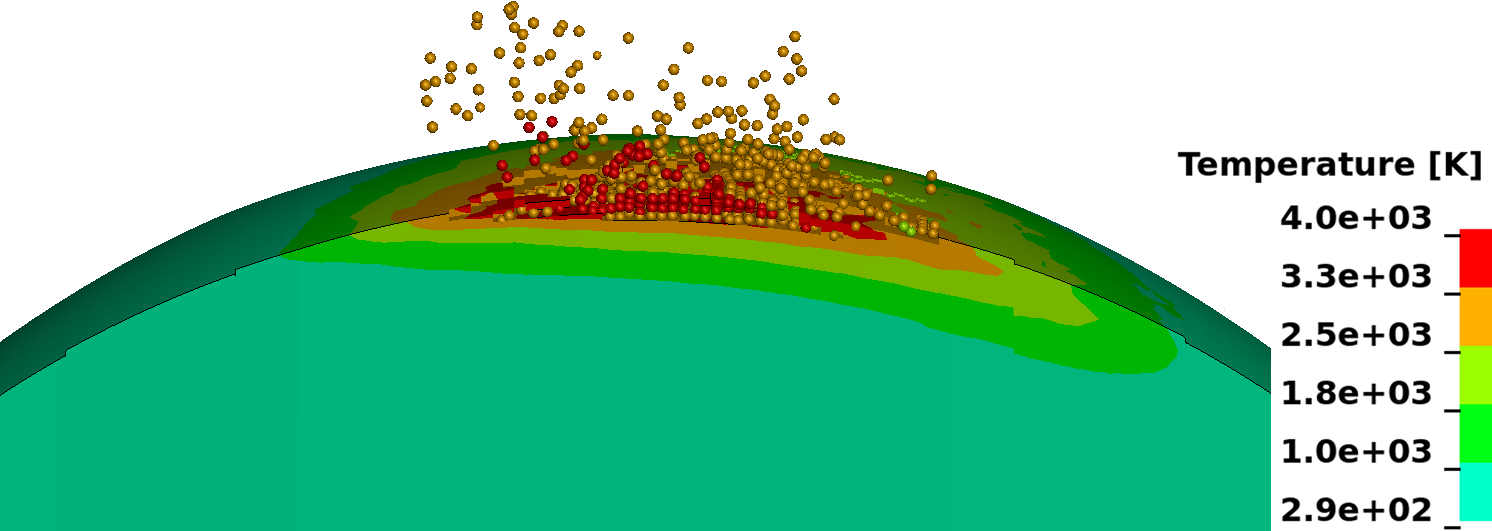}
          \put(2,28){\hbox{\kern3pt\textcolor{black}{\textbf{c)} }}}
          \put(59,28){\hbox{\kern3pt\textcolor{black}{\textbf{t=1.0 ms} }}}
          \end{overpic}
        }\\
        \subfloat{%
          \begin{overpic}[width = 2.15in]{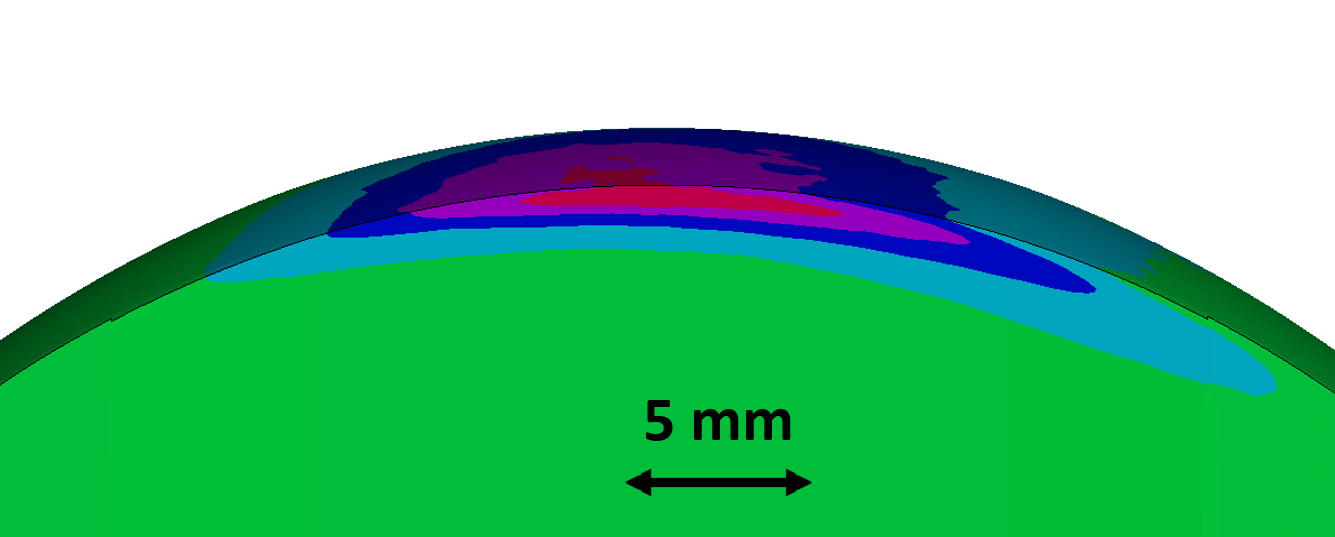}
          \put(2,35){\hbox{\kern3pt\textcolor{black}{\textbf{d)} }}}
          \end{overpic}
        }
        \hspace{-3mm}
        \subfloat{%
          \begin{overpic}[width = 2.2in]{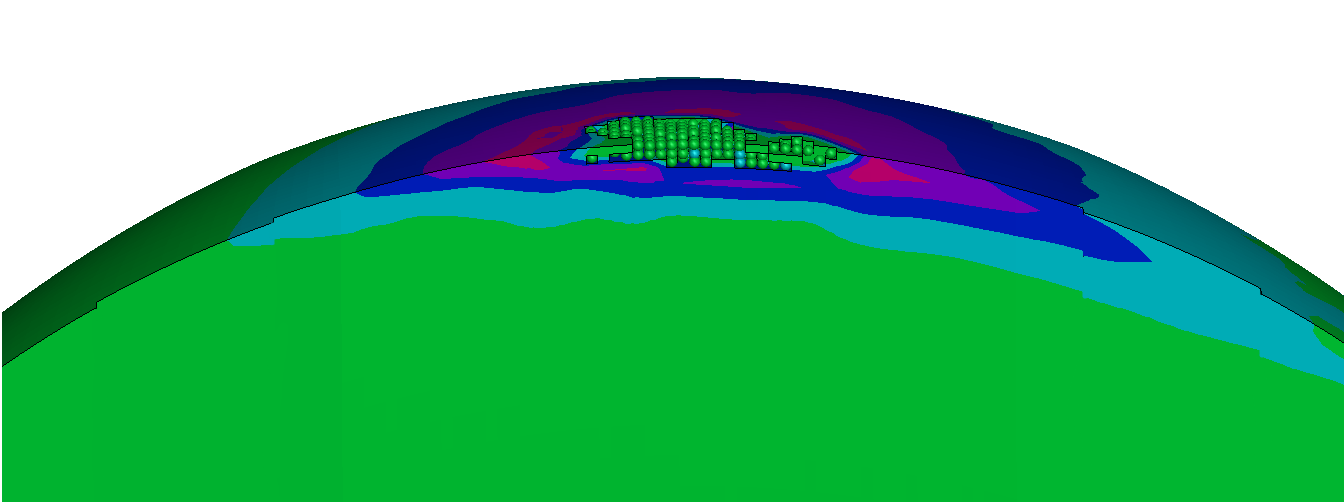}
          \put(2,35){\hbox{\kern3pt\textcolor{black}{\textbf{e)} }}}
          \end{overpic}
        }
        \hspace{-3mm}
        \subfloat{%
          \begin{overpic}[width = 2.7in]{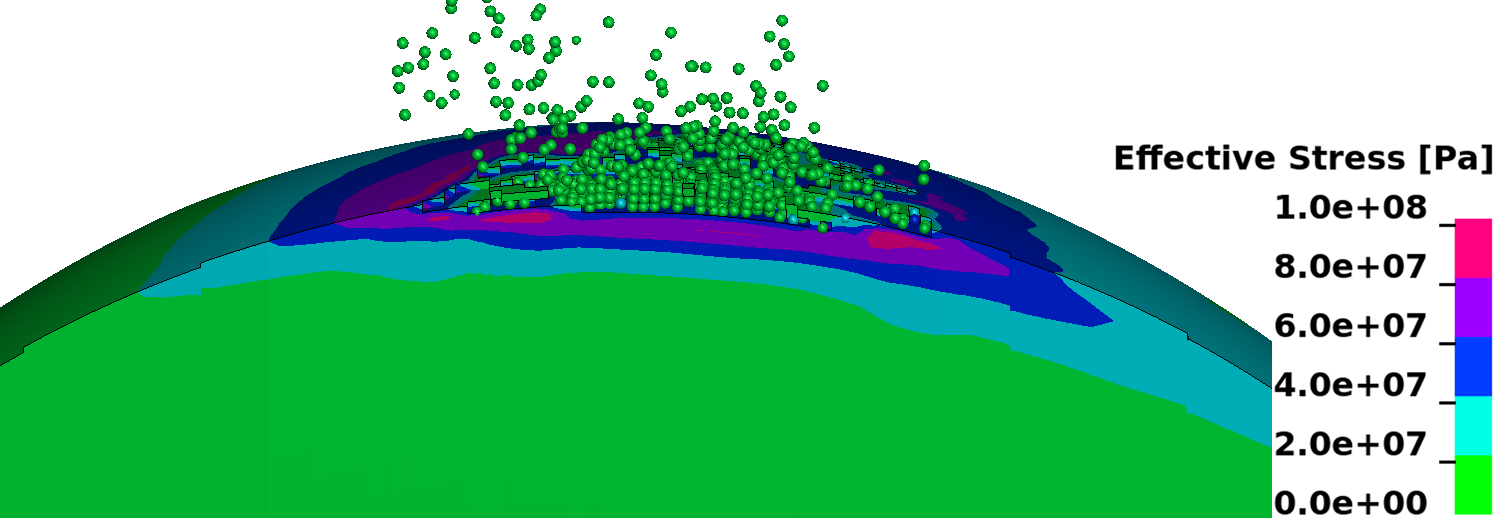}
          \put(2,27){\hbox{\kern3pt\textcolor{black}{\textbf{f)} }}}
          \end{overpic}
        }\\
        \subfloat{%
          \begin{overpic}[width = 2.15in]{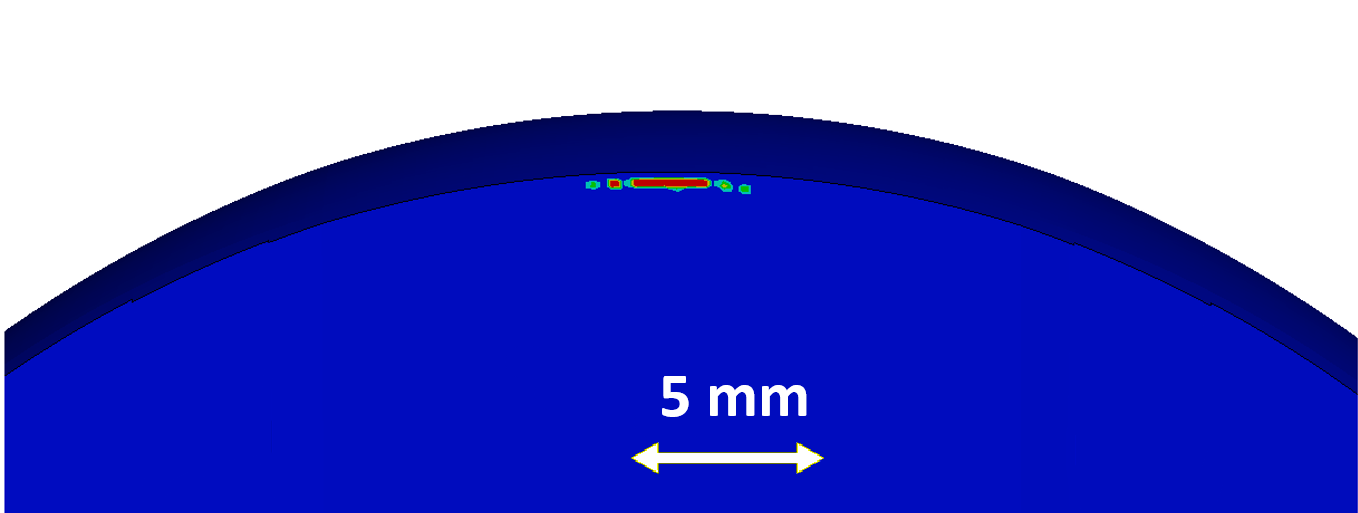}
          \put(2,35){\hbox{\kern3pt\textcolor{black}{\textbf{g)} }}}
          \end{overpic}
        }
        \hspace{-3mm}
        \subfloat{%
          \begin{overpic}[width = 2.2in]{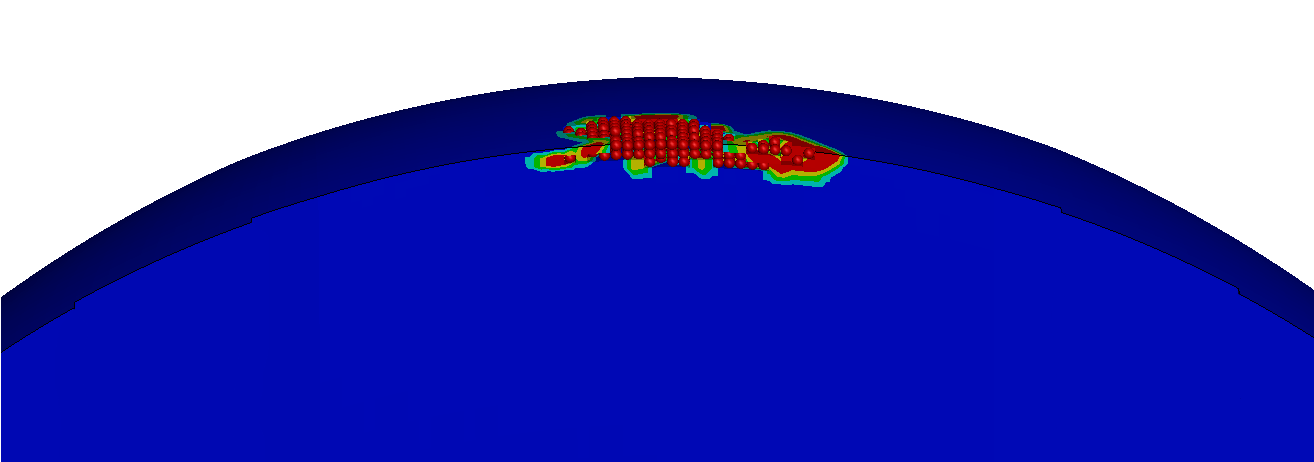}
          \put(2,35){\hbox{\kern3pt\textcolor{black}{\textbf{h)} }}}
          \end{overpic}
        }
        \hspace{-3mm}
         \subfloat{%
          \begin{overpic}[width = 2.67in]{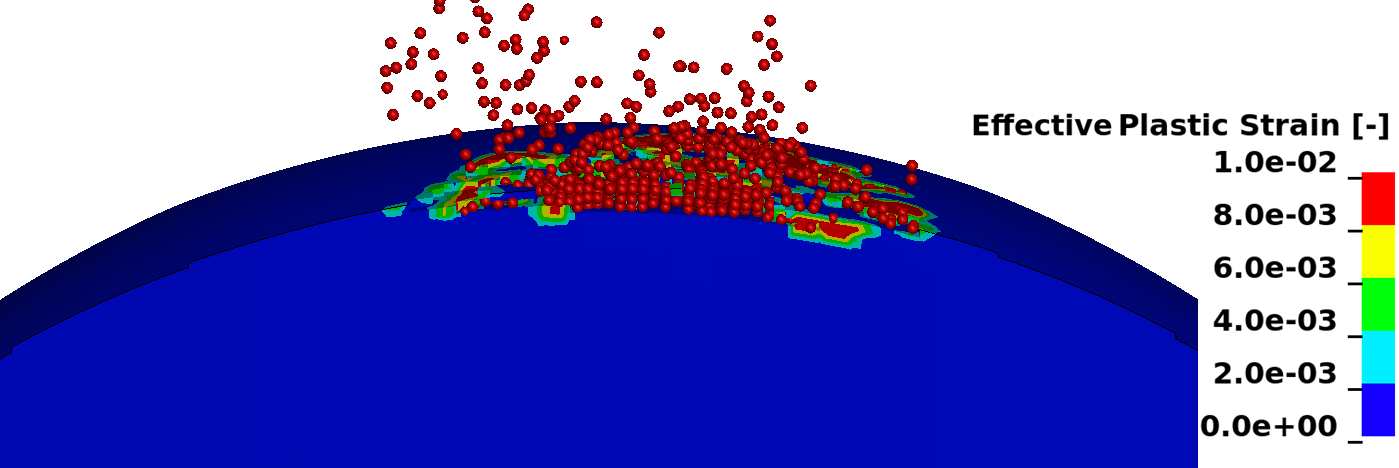}
          \put(2,27){\hbox{\kern3pt\textcolor{black}{\textbf{i)} }}}
          \end{overpic}
        }\\
    \end{tabular}
    \caption{LS-DYNA coupled FEM-SPH simulation results for Case 1 (\#191366): enlarged view of the damaged region (see the scale bar for the dimensions). The $y=0$ cross-sections of the 3D computational domain are shown for the temperature (a--c), the von Mises stress (d--f), and the effective plastic strain (g--i) at $t = 0.74$\,ms (first column), $t = 0.80$\,ms (second column), and $t = 1.00$\,ms (third column). Note that, since $\epsilon_{\mathrm{f}}=0.025$, the color bar in panel (g--i) is saturated at 0.01 to enable visualization.}
    \label{fig:results_panel}
\end{figure*}

\subsection{Onset of failure with different material models}

Despite the adoption of a different constitutive equation and a different failure criterion, the present LS-DYNA simulations capture the same mechanism of material fragmentation as with our earlier COMSOL Multiphysics simulations\,\cite{Ratynskaia_2025}. To be more specific, the RE energy deposition profile peaks beneath the surface, as reflected in the temperature profiles. This results in the development of strong compressive stresses beneath the surface, which ultimately drive material failure and the subsequent expulsion of debris\,\cite{Ratynskaia_2025,De_Angeli_2023}.

This temporal evolution is illustrated in Fig.\ref{fig:results_panel}, where spatial maps of the temperature, the equivalent stress, and the effective plastic strain are shown for a specific cross-section of the three-dimensional domain at three different time instances. In the first column, these fields are illustrated at $t = 0.74$\,ms -- only a few time steps before the first FEM element fails. Fig.\ref{fig:results_panel}(a) reveals a non-monotonic temperature profile, similar to Fig.3 of Ref.\cite{Ratynskaia_2025}. Fig.\ref{fig:results_panel}(d) shows a large stress accumulation just beneath the surface, where the plastic strain begins to increase, as shown in Fig.\ref{fig:results_panel}(g).

At $t = 0.80$\,ms -- a few time steps after the first FEM element fails, several elements have already been replaced by SPH particles. As described in Sec.~\ref{sec:modeling}.4, the SPH nodes retain a constant temperature equal to that inherited from the replaced element. Once they stop interacting with surrounding FEM elements, their stress tensor is set to zero, consistent with the assumption that they represent fragmented material and thus cannot sustain deviatoric stresses. As evident from Fig.\ref{fig:results_panel}(h), the effective plastic strain continues to increase in the damaged region, while all SPH nodes are characterized by an EPS that exceeds $2.5\%$.

Finally, Figs.\ref{fig:results_panel}(c,f,i) correspond to the end of the RE loading, at $t = 1$\,ms. Then, the failed region is already close to its final extension, while newly failed elements that have converted into SPH particles are visible as ejecta, having detached from the surface and propagating outwards. This is consistent with the camera ($\sim1$\,ms frame rate) observations where debris release is detected about 1 ms after the RE impact \cite{Hollmann2025}.


\subsection{Fragmentation and total eroded volume}

The onset of fragmentation predicted by thermoelastic-plastic equations coupled with the EPS criterion occurs at $t = 0.76$\,ms, which is reasonably close to the $0.6$\,ms predicted by purely thermoelastic equations coupled with the Rankine criterion\,\cite{Ratynskaia_2025}. Most material erosion takes place within the loading time of 1\,ms. However, stress relaxation, interactions between FEM elements and SPH–FEM interactions, trigger additional failure post-loading. The  SPH–FEM interactions are handled via a penalty-based contact algorithm that enforces non-penetration.

\begin{figure}[!h]
\centering
\begin{subfigure}{0.235\textwidth}
    \centering
    \includegraphics[width=\linewidth]{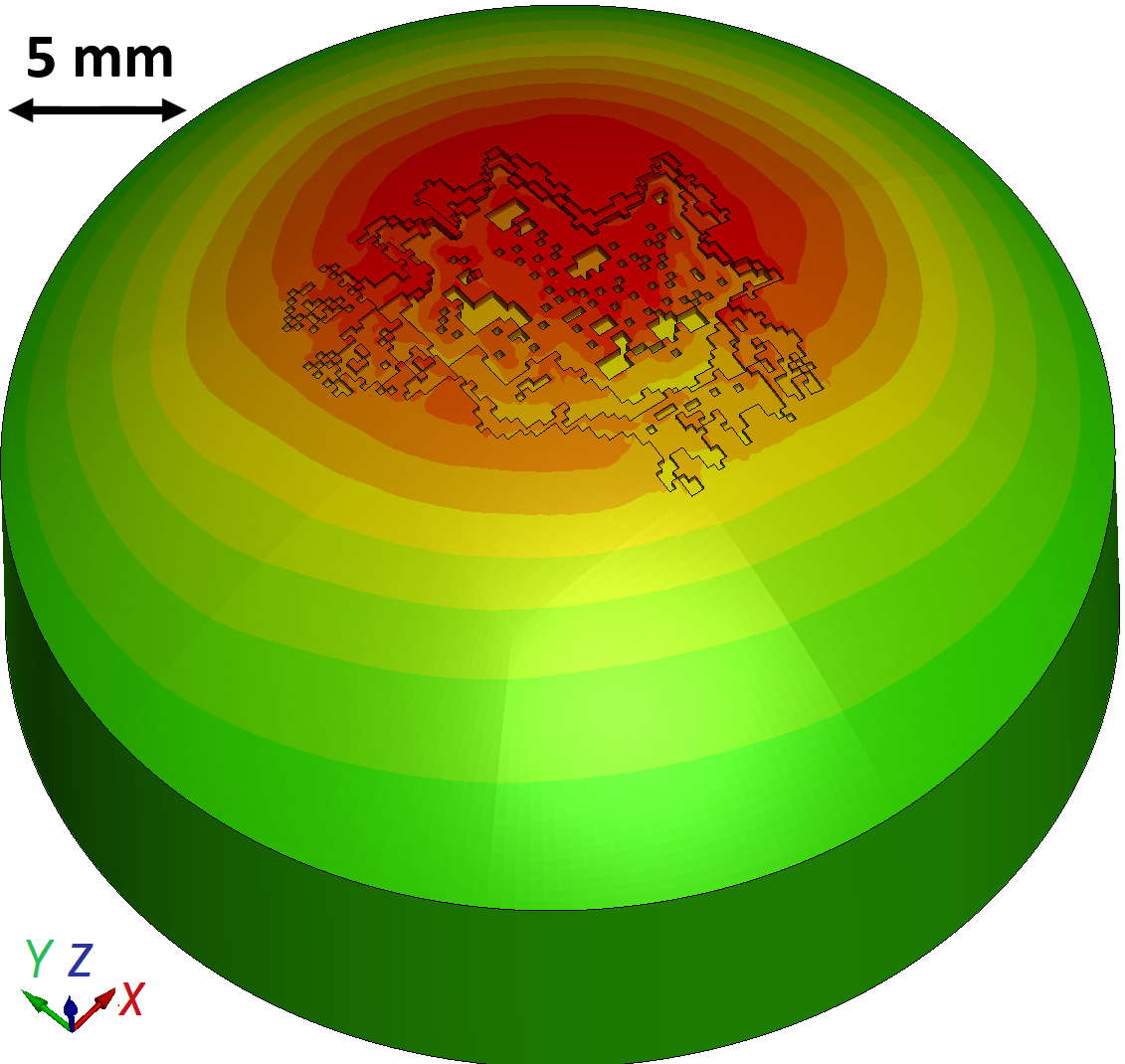}
    \caption{}
\end{subfigure}
\hspace{0.0001\textwidth}
\begin{subfigure}{0.235\textwidth}
    \centering
    \includegraphics[width=\linewidth]{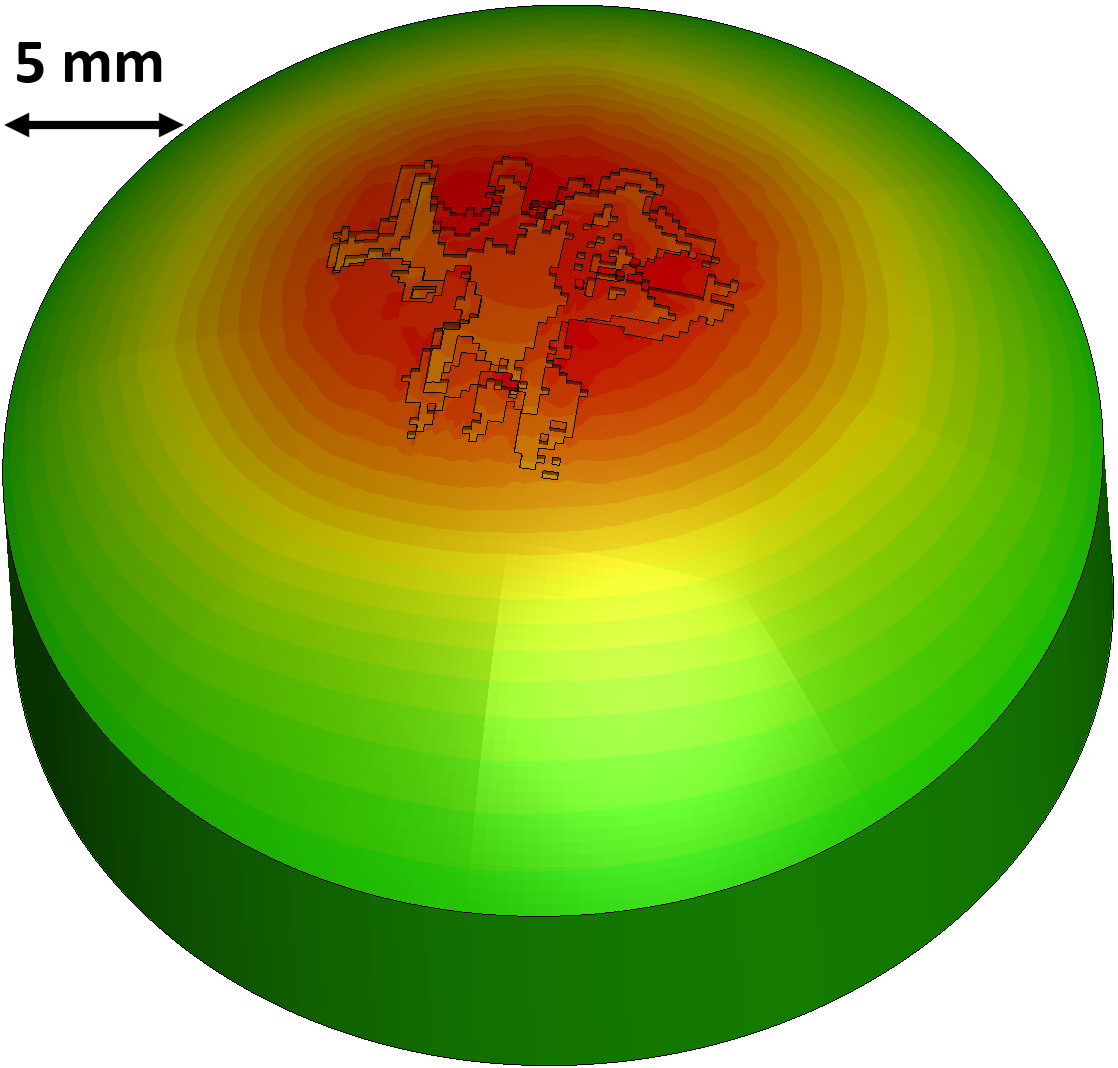}
    \caption{}
\end{subfigure}
\caption{Final RE-induced graphite damage at $t=4$\,ms; (a) Case 1, \#191366, (b) Case 2, \#200241. The colormap indicates element displacement.}
\label{fig:damage}
\end{figure}

The simulation results for the final damage, acquired at $t = 4$\,ms, are in good agreement with the experimental results of Table \ref{tab:summary_exposure}. In particular, a total eroded volume of $\sim130$\,mm$^3$ is found. Moreover, the topology of the damaged region, shown in Fig.~\ref{fig:damage} (a), is characterized by a maximum depth of $\sim1.3$\,mm and a surface area of $\sim120$\,mm$^2$.

The simulation results for the failed region also feature several macro-cracks that are developed in the peripheral area. These features are not observed in the ATJ graphite sample, which has a more regular circular damage pattern, albeit with a relatively high surface roughness. The cracks remain superficial and may be attributed to the stress redistribution occurring after the FEM element failure. A fully self-consistent modeling of crack propagation would require detailed knowledge of the sample's defect distribution and grain structure\,\cite{Levy_2010,Shenoy_2003}; it lies beyond the scope of this work.


\subsection{Debris}

\begin{figure}
    \centering
    \includegraphics[width=0.86\linewidth]{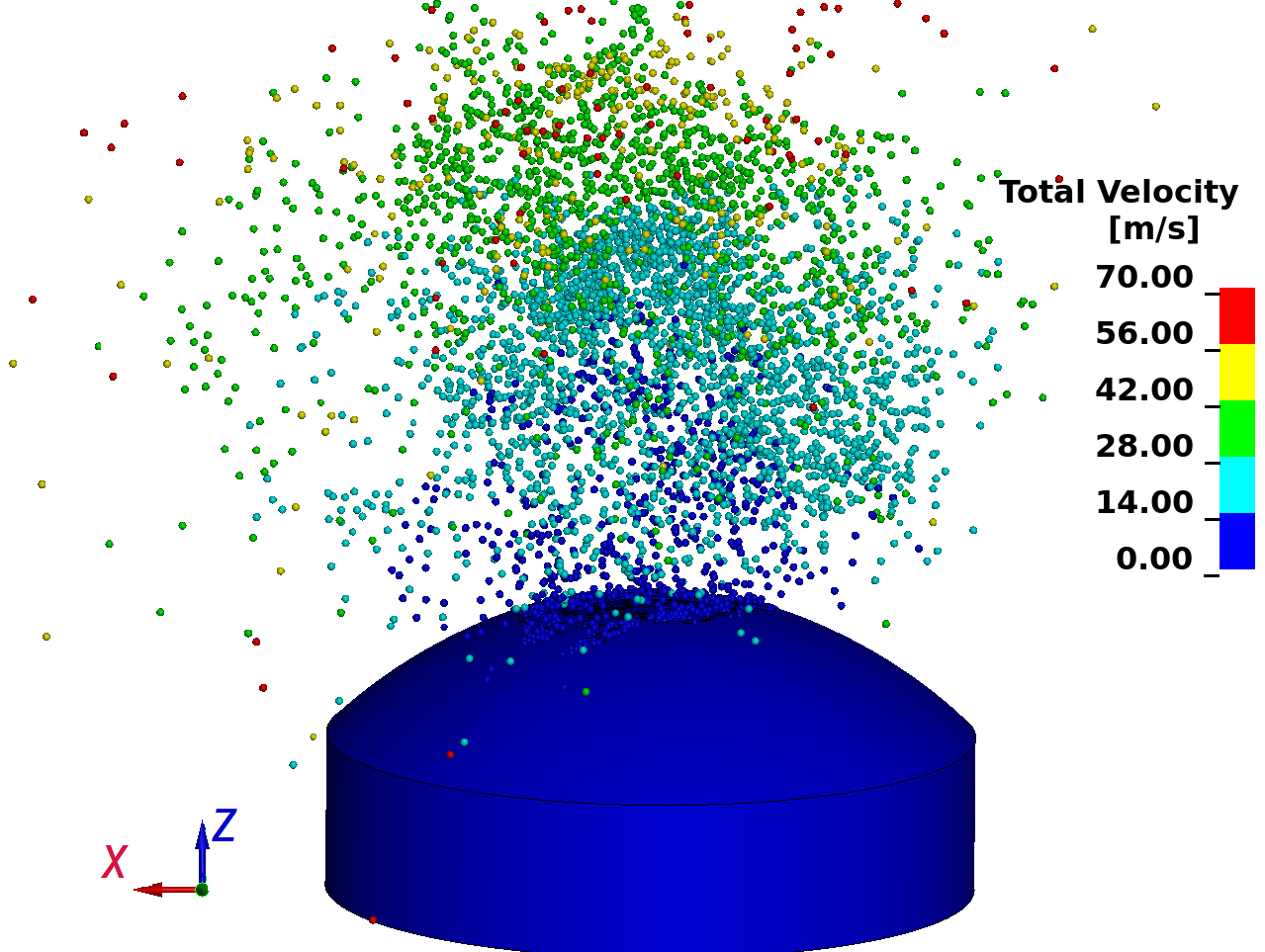}
    \caption{Sample fragmentation in Case 1 (\#191366) at $t=2$\,ms focusing on the debris released. The color coding represents the speed. SPH particles are represented as spheres for visualization purposes only.}
    \label{fig:debris_prop}
\end{figure}

The hybrid FEM-to-SPH approach allows the tracking of fragmented material after the explosion occurs. A snapshot of the released debris, captured at $t = 2$\,ms, is shown in Fig.\ref{fig:debris_prop}. The SPH particles are represented as spheres of constant radius for visualization purposes only; their size and shape does not correspond to the actual fragment. The color bar indicates their speed. Fig.\ref{fig:debris_prop} reveals that the debris direction is concentrated predominantly towards the surface normal of the top surface and that the ejected dust is radially diffusing outwards in agreement with the camera observations. Fig.\ref{fig:histogram_velocities}(a) features a comparison between the SPH and tracked debris speed distribution, with mean values of 33\,m/s and 46\,m/s, respectively. 




\section{Results for Case 2 (\#200241)}\label{sec:results2}

\subsection{Fragmentation and total eroded volume}

In Case 2, the RE deposited total energy was comparable to that of Case 1, but the removed volume was about a factor of 5 lower. This discrepancy can be attributed to the energy deposition map characteristics discussed in Sec.~\ref{sec:modeling}. In particular, the larger wetted area, smoother spatial gradients and lower peak values of the energy density, lead to a weaker and delayed growth of the strain field. Consequently, a considerably smaller fraction of the material volume satisfies the failure criterion.

The simulations predict the onset of fragmentation at $t=0.91$\,ms, \emph{i.e.,} delayed relative to Case 1. Again, this is consistent with the camera observations ($\sim1$\,ms frame rate) revealing debris ejection at about 1\,ms after the RE incidence. Case 2 has also been earlier simulated within linear thermo-elasticity theory combined with Rankine’s failure theory\,\cite{Hollmann_2025b}, leading to a $t=0.9$\,ms prediction for the brittle failure onset, which is in full agreement with the present predictions. The resulting final damage topology is illustrated in Fig.~\ref{fig:damage}(b). The total failed volume is $\sim60$\,mm$^3$ with $\sim1$\,mm maximum depth and $\sim70$\,mm$^2$ surface area, to be compared to the measurements reported in Table~\ref{tab:summary_exposure}. 


\subsection{Debris}

\begin{figure}
\centering
\begin{subfigure}{0.239\textwidth}
    \centering
    \includegraphics[width=\linewidth]{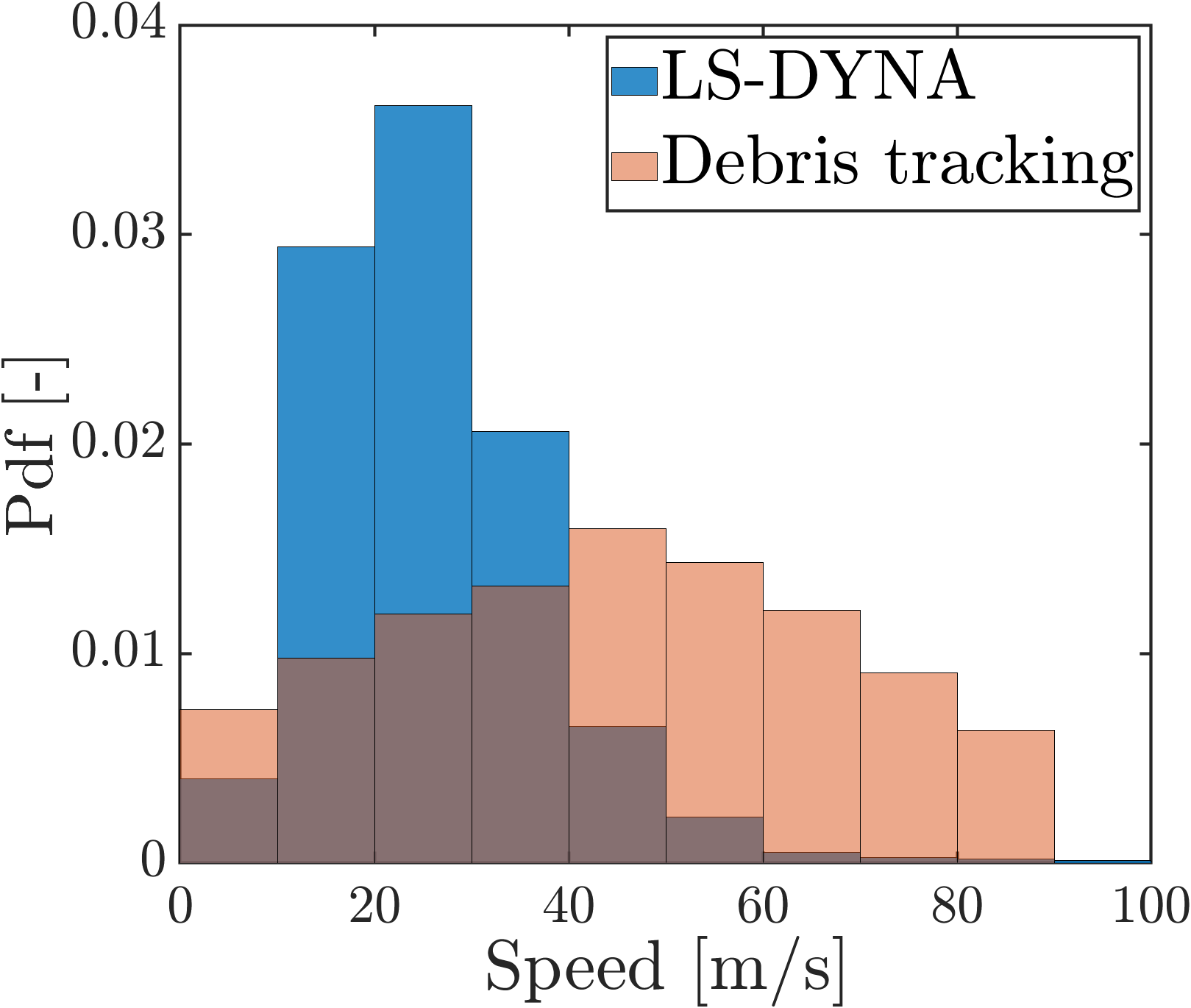}
    \caption{}
\end{subfigure}
\hspace{0.001\textwidth}
\begin{subfigure}{0.23\textwidth}
    \centering
    \includegraphics[width=\linewidth]{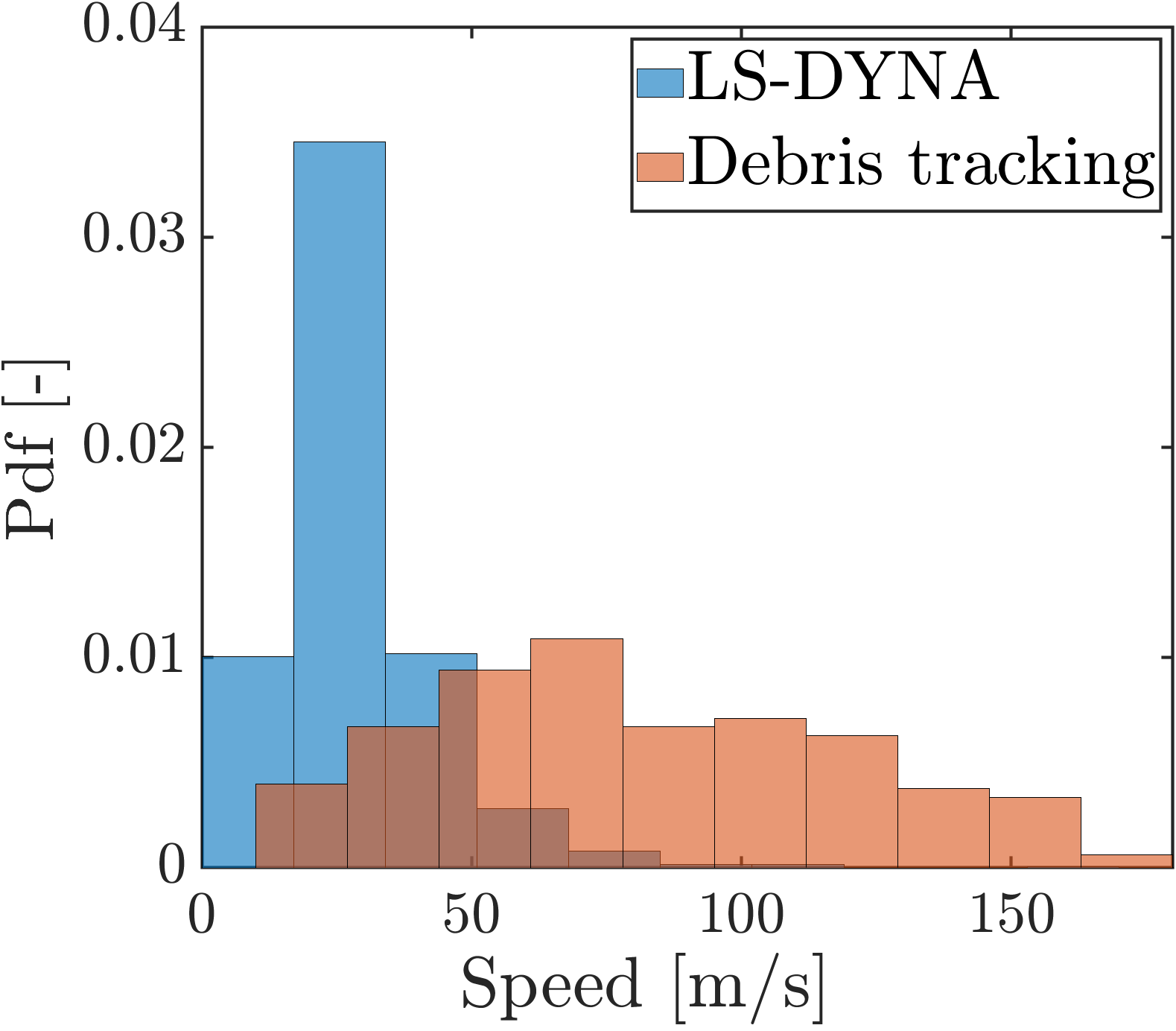}
    \caption{}
\end{subfigure}
\caption{ The SPH particle (blue) vs the experimentally reconstructed debris (red) speed distributions; (a) Case 1 - \#191366, (b) Case 2 - \#200241.}
\label{fig:histogram_velocities}
\end{figure}

Fig.\ref{fig:histogram_velocities}(b) compares the experimentally reconstructed and SPH particle speed distributions. Recall that in Case 2, the tracking results have higher uncertainties due to the lower video quality, see Sec.~\ref{sec:experimental}. Overall, this exposure is characterized by higher speeds, compare (a) and (b).


\section{Critical assessment and conclusions}

A complete model of the coupled thermomechanical response of brittle sublimating materials to volumetric RE heat loads has been developed. The combination of finite element analysis with smoothed-particle hydrodynamics allows for the description of the entire range of deformations for the first time. The work-flow begins with the RE energy deposition (with the reconstructed RE impact parameters constituting a critical input), proceeds with the linear thermomechanical response up to the onset of failure and culminates in the nonlinear stage of material fragmentation and debris expulsion, as observed in RE-driven explosive events in tokamaks. This crucial step in RE-induced damage modeling has been facilitated by a series of specially designed controlled experiments in the DIII-D tokamak\,\cite{Hollmann2025,Hollmann_2025b}, recent advances in the understanding of RE dynamics\,\cite{Ratynskaia_2025b,Breizman2019} and progress in the understanding of material damage induced by fast transients\,\cite{Krieger_2025,Ratynskaia2020}.

The developed work-flow is shown to be capable of robust predictions of brittle material explosions under RE impacts in terms of the eroded region and the blown-off volume. It also provides quantitative results on the debris, despite the emergence of cracking that has not been observed in the experiments. Since the cracks are mainly superficial, they have an insignificant effect on the eroded volume and the debris expulsion. Thus, further physics and numerical efforts in minimizing cracking are rather unnecessary. Finally, the comparison between the debris speed distributions must be viewed from two perspectives. With respect to the simulations, the mean speed underestimation is intrinsic to the adaptive FEM-to-SPH approach under subsurface initiated failure. SPH particles are initially generated inside the bulk, surrounded by intact loadbearing finite element neighbors under near-symmetric contact forces, while the plasticity admitted by JH-2 channels their inherited elastic energy mainly into plastic work rather than translational motion. The fragments that eventually reach a free surface acquire the expected speeds, while particles born deep underneath mainly populate the low-speed part of the distribution. With respect to the experiments, it must be stressed that the reconstructed debris speed distributions suffer from large uncertainties, that are imposed by limited camera resolution and exacerbated by the strong thermal radiation from hot debris and by the high concentration of fragments. 

For the RE-driven explosion of a brittle, sublimating, light and low-Z material such as graphite, the SPH approach is a valuable tool that enables simulations of debris expulsion and provides an estimate of the debris speed. Such estimates serve to demonstrate that the numerical approach is adequate to capture the principal  characteristics of the dedicated experiments, but for the fusion reactor relevant case of the RE-driven explosion of tungsten, a more quantitative and detailed knowledge of the debris size and speed is required. This is due to the fact that ejected fragments, if mobilized in subsequent discharges, would become a high Z impurity source, thus their sizes need to be quantified\,\cite{Ratynskaia_2022R}. Moreover, fast, heavy metal debris can lead to non-localized damage in terms of cratering upon their impacts on the surrounding in-vessel components\,\cite{Tolias_2023,DeAngeli_2024,Tolias_2026}. Other mesh-free techniques may need to be considered for the modeling of tungsten explosions, see Ref.\cite{Ratynskaia_2025b} for a more detailed discussion.


\section*{Acknowledgments}

\noindent SR acknowledges the financial support of the Swedish Research Council under Grant No.2025-05867. The work has been performed within the framework of the EUROfusion Consortium,\,funded by the European Union via the Euratom Research and Training Programme (Grant Agreement No\,101052200 - EUROfusion). The views and opinions expressed are however those of the authors only and do not necessarily reflect those of the European Union, the European Commission or the ITER Organization. The European Union, European Commission or ITER Organization cannot be held responsible for them. The simulations were enabled by resources provided by the National Academic Infrastructure for Supercomputing in Sweden (NAISS) at the NSC (Link\"oping University) partially funded by the Swedish Research Council through grant agreement No\,2022-06725.  The HMU authors acknowledge the support with computational time granted by the Greek Research \& Technology Network (GRNET) in the National HPC facility ARIS—under Project No.\,pr018017-LaMPIOS IV. This work was performed under the auspices of the U.S. Department of Energy by Lawrence Livermore National Laboratory under Contract DE-AC52-07NA27344. This material is based upon work supported by the U.S. Department of Energy, Office of Science, Office of Fusion Energy Sciences, using the DIII-D National Fusion Facility, a DOE Office of Science user facility, under Award(s) DE-FC02-04ER54698.  This work was supported in part by the U.S. Department of Energy under DE-FG02-07ER54917. This material is based upon work supported by the U.S. Department of Energy, Office of Science, Office of Fusion Energy Sciences, under Contract Nos. DE-AC05-00OR22725. This research also uses resources of the National Energy Research Scientific Computing Center (NERSC), a U.S. Department of Energy Office of Science User Facility operated under Contract No. DE-AC02-05CH11231. This report was prepared as an account of work sponsored by an agency of the United States Government. Neither the United States Government nor any agency thereof, nor any of their employees, makes any warranty, express or implied, or assumes any legal liability or responsibility for the accuracy, completeness, or usefulness of any information, apparatus, product, or process disclosed, or represents that its use would not infringe privately owned rights. Reference herein to any specific commercial product, process, or service by trade name, trademark, manufacturer, or otherwise does not necessarily constitute or imply its endorsement, recommendation, or favoring by the United States Government or any agency thereof. The views and opinions of authors expressed herein do not necessarily state or reflect those of the United States Government or any agency thereof.

\section*{References}
\noindent  
\bibliography{biblio_new}

\end{document}